\documentclass[aps,superscriptaddress,preprint,pra,epsfig,showpacs,amsmath,amsfonts,amssymb,floatfix]{revtex4-1}
\usepackage{amsmath,amsfonts,amssymb,color}
\usepackage{amsthm}
\usepackage{leftidx}
\usepackage{graphicx}
\usepackage{xcolor}
\usepackage{dcolumn}
\usepackage{bm}
\usepackage{epstopdf}
\usepackage{epsfig}
\usepackage{environ}
\usepackage{pdfcomment}

\usepackage{float}
\usepackage[T1]{fontenc}
\usepackage[latin9]{inputenc}
\usepackage{setspace}
\usepackage{esint}

\usepackage{geometry}
\usepackage{mathrsfs}
\usepackage{amsmath}
\usepackage{amssymb}
\usepackage{graphicx}
\usepackage{setspace}

\begin{document}
	

\title{Non-Hermitian Floquet phases with even-integer topological invariants
	in a periodically quenched two-leg ladder}

\author{Longwen Zhou}
\email{zhoulw13@u.nus.edu}
\affiliation{%
	Department of Physics, College of Information Science and Engineering, Ocean University of China, Qingdao, China 266100
}

\date{\today}




\begin{abstract}
Periodically driven non-Hermitian systems could possess exotic nonequilibrium
	phases with unique topological, dynamical and transport properties. In this work, we introduce an experimentally realizable two-leg ladder
	model subjecting to both time-periodic quenches and non-Hermitian
	effects, which belongs to an extended CII symmetry class. Due to the
	interplay between drivings and nonreciprocity, rich non-Hermitian
	Floquet topological phases emerge in the system, with each of them
	been characterized by a pair of even-integer topological invariants
	$(w_{0},w_{\pi})\in2\mathbb{Z}\times2\mathbb{Z}$. Under the open
	boundary condition, these invariants further predict the number of
	zero- and $\pi$-quasienergy modes localized around the edges of the
	system. We finally construct a generalized version of the mean chiral
	displacement, which could be employed as a dynamical probe to the
	topological invariants of non-Hermitian Floquet phases in the CII
	symmetry class. Our work thus introduces a new type of non-Hermitian
	Floquet topological matter, and further reveals the richness of topology
	and dynamics in driven open systems.
\end{abstract}

\maketitle
\section{Introduction\label{sec:Intro}}

Non-Hermitian states of matter have attracted great attention in recent
years due to their intriguing dynamical and topological properties
(see \cite{NHRev1,NHRev2,NHRev3,NHRev4,NHRev5,NHRev6,NHRev7} for reviews). Theoretically, a wide range of non-Hermitian topological
phases and phenomena have been classified and characterized
according to their symmetries \cite{Class1,Class2,Class3,Class4,Class5,Class6,Class7} and dynamical signatures \cite{Dym1,Dym2,Dym3,Dym4,Dym5,Dym6}.
Experimentally, non-Hermitian topological matter have also been realized
in cold atom \cite{ColdAtom1,ColdAtom2}, photonic \cite{Photonic1,Photonic2,Photonic3,Photonic4}, acoustic \cite{Acoustic1,Acoustic2,Acoustic3}, electrical circuit \cite{Circuit1,Circuit2,Circuit3} systems and nitrogen-vacancy-center in diamond \cite{NVCent1}, leading to potential applications such as topological lasers \cite{TopoLas1,TopoLas2,TopoLas3} and high-performance sensors \cite{EPSense1,EPSense2,EPSense3,EPSense4}.

Recently, the study of non-Hermitian physics has been extended to
Floquet systems, in which the interplay between time-periodic driving
fields and gains/losses or nonreciprocal effects could potentially
yield topological phases that are unique to driven non-Hermitian systems \cite{ZhouNHFTP1,ZhouNHFTP2,ZhouNHFTP3,ZhouNHFTP4,ZhouNHFTP5,NHFTP1,NHFTP2,NHFTP3,NHFTP4,NHFTP5,NHFTP6,NHFTP7,NHFTP8,NHFTP9,NHFTP10}.
In early studies, various non-Hermitian Floquet topological phases and phenomena
have been discovered, including non-Hermitian
Floquet topological insulators \cite{ZhouNHFTP1,ZhouNHFTP2,ZhouNHFTP5,NHFTP1,NHFTP2}, superconductors \cite{ZhouNHFTP4}, semimetals \cite{NHFTP10}, and skin
effects \cite{NHFTP3,NHFTP4}. Meanwhile, the time-averaged spin texture and mean
chiral displacement have been suggested as two dynamical tools to
extract the topological invariants of non-Hermitian Floquet systems \cite{ZhouNHFTP1,ZhouNHFTP2,ZhouNHFTP3,ZhouNHFTP5}.
These discoveries extend the boundary of nonequilibrium phases of
matter to driven non-Hermitian systems, and shed light on new approaches
for the detection of their intriguing features.

In previous studies, non-Hermitian Floquet phases were explored mainly in two-band systems.
In this work, we uncover a new type of non-Hermitian Floquet topological
matter in the extended CII symmetry class, which contains at least four quasienergy bands. The system can be realized in
a periodically quenched nonreciprocal two-leg ladder, as introduced
in Sec. \ref{sec:Model}. Each topological phase of the system is characterized
by a pair of even-integer winding numbers $(w_{0},w_{\pi})\in2\mathbb{Z}\times2\mathbb{Z}$.
With the change of the nonreciprocal parameters of the model, we find
rich non-Hermitian Floquet phases with large winding numbers, and
various topological phase transitions induced by non-Hermitian effects,
as presented in Sec. \ref{sec:WN}. In Sec. \ref{sec:BBC}, we obtain
multiple quartets of non-Hermitian Floquet edge modes in our system
at zero and $\pi$ quasienergies under the open boundary condition
(OBC), and establish the correspondence between the number of these
modes and bulk topological invariants $(w_{0},w_{\pi})$. In Sec.
\ref{sec:MCD}, we explore the dynamical aspects of our model by generalizing
the mean chiral displacement (MCD) to non-Hermitian Floquet systems
in the CII symmetry class, and demonstrate the relationship between
the MCDs and the topological winding numbers $(w_{0},w_{\pi})$. Finally,
we summarize our findings and discuss potential future directions
in Sec. \ref{sec:Summary}.

\section{Model and symmetry\label{sec:Model}}


The model we are going to investigate can be viewed as a driven, non-Hermitian
version of the Creutz ladder (CL) with spin-$1/2$ fermions and spin-orbit
couplings (or spinless particles with four sublattice degrees of freedom).
The CL model refers to a quasi-one-dimensional lattice formed by two
coupled legs and subjected to a perpendicular magnetic flux~\cite{CL0}.
It could possess symmetry-protected degenerate zero modes at its boundaries,
and therefore belong to one of the earliest examples of a topological
insulator~\cite{CL0}. In later studies, the CL model has be realized in
photonic \cite{CLExp1,CLExp2} and cold atom \cite{CLExp3,CLExp4} systems, and utilized in the investigations
of Aharonov-Bohm cages \cite{CL1,CL2}, topological pumping \cite{CL3}, localization \cite{CL4,CL5} and many-body
topological matter \cite{CL6,CL7,CL8,CL9,CL10}. Recently, spin-$1/2$ extensions of the CL model
have also been explored in several studies \cite{SCL1,SCL2,SCL3}, leading to the
discoveries of richer topological features. Furthermore, when time-periodic
drivings are applied to the spin-$1/2$ CL, a series of Hermitian
Floquet topological phases in the CII symmetry class were found \cite{FSCL}. Each of these phases is characterized by a pair of even-integer
topological winding numbers, quantized dynamics of bulk states, together
with degenerate quartets of zero and $\pi$ Floquet edge modes under
the OBC \cite{FSCL}. In this work, the construction of our system can
be viewed as a non-Hermitian extension of the model studied in Ref.~\cite{FSCL}, and will be referred to as the non-Hermitian periodically quenched
two-leg ladder (PQTLL).

The Hamiltonian of the non-Hermitian PQTLL model takes the form
\begin{equation}
H(t)=f(t)H_{\parallel}+g(t)H_{\bot},\label{eq:Ht}
\end{equation}
where

\begin{equation}
f(t)\equiv \sum_{j\in\mathbb{Z}}[\theta(j+1/2-t)-\theta(j-t)],\quad g(t)\equiv \sum_{j\in\mathbb{Z}}[\theta(j+1-t)-\theta(j+1/2-t)], \label{eq:fgt}
\end{equation}
describe the quench protocols in the first and second halves of each
driving period $T$, with $\theta(t)$ being the step function. Throughout this work, we set $\hbar=T=1$ as the convention of units.
In the lattice representation, the Hamiltonian components $H_{\parallel}$ and
$H_{\bot}$ are explicitly given by

\begin{alignat}{1}
H_{\parallel}& = \sum_{n}J_{x}(|n\rangle\langle n+1|+{\rm H.c.})\sigma_{0}\otimes\tau_{z} - \sum_{n}iV(|n\rangle\langle n+1|-{\rm H.c.})\sigma_{y}\otimes\tau_{0},\label{eq:Hpara}\\
H_{\bot}& = \sum_{n}J_{y}|n\rangle\langle n|\sigma_{0}\otimes\tau_{x} + \sum_{n}iJ_{d}(|n\rangle\langle n+1|-{\rm H.c.})\sigma_{z}\otimes\tau_{x}.\label{eq:Hbot}
\end{alignat}
Here $n=1,...,N$ are the indices of unit cells, which are arranged along the
horizontal $(x)$ direction of the ladder. $\sigma_{0}$ and $\tau_{0}$
are both $2\times2$ identity matrices. Each unit cell of the ladder
contains two spin and sublattice components, and $\sigma_{x,y,z}$,
$\tau_{x,y,z}$ are Pauli matrices acting on the spin-$1/2$ and sublattice
degrees of freedom, respectively. The system parameters $J_{x}$ and $J_{y}$ represent
the intercell and intracell hopping amplitudes along the horizontal
($x$) and vertical ($y$) directions of the ladder. $J_{d}$ controls
the coupling strength between different sublattices in adjacent unit
cells, and $V$ describes the amplitude of spin-orbit coupling among
particles with opposite spins in the same sublattice of nearest-neighbor
unit cells. In this work, we allow $J_{y}$ and $J_{d}$ to take complex
values, i.e., $J_{y}=J_{y}^{r}+{\rm i}J_{y}^{i}$ and $J_{d}=J_{d}^{r}+{\rm i}J_{d}^{i}$.
Physically, the imaginary parts $J_{y}^{i}$ and $J_{d}^{i}$ could
describe the nonreciprocity of hoppings along the vertical and diagonal
directions of the ladder.

The Floquet operator of the non-Hermitian PQTLL model, which corresponds
to its evolution operator over a complete driving period (e.g., from
$t=j+0^{-}$ to $j+1+0^{-}$), can be expressed as

\begin{equation}
U=\mathscr{T}e^{-{\rm i}\int_{0}^{1}H(t)dt}=e^{-\frac{{\rm i}}{2}H_{\bot}}e^{-\frac{{\rm i}}{2}H_{\parallel}},\label{eq:U}
\end{equation}
where $\mathscr{T}$ is the time-ordering operator. The quasienergy
spectrum $\varepsilon$ of the system can be obtained by solving the eigenvalue
equation $U|\psi\rangle=e^{-{\rm i}\varepsilon}|\psi\rangle$ under
a fixed boundary condition, where $|\psi\rangle$ is the corresponding
right eigenvector of $U$. With a ladder of $N$ unit cells and under
the period boundary condition (PBC), one can perform the Fourier transform
$|n\rangle=\frac{1}{\sqrt{N}}\sum_{k}e^{-{\rm i}nk}|k\rangle$ to
express $U$ in momentum space as $U=\sum_{k}|k\rangle U(k)\langle k|$.
Here $k\in[-\pi,\pi)$ is the quasimomentum, and

\begin{alignat}{1}
U(k)& = e^{-{\rm i}h_{\bot}(k)}e^{-{\rm i}h_{\parallel}(k)},\label{eq:Uk}\\
h_{\parallel}(k)& = J_{x}\cos k\sigma_{0}\otimes\tau_{z}+V\sin k\sigma_{y}\otimes\tau_{0},\label{eq:hparak}\\
h_{\bot}(k)& = \frac{J_{y}}{2}\sigma_{0}\otimes\tau_{x}-J_{d}\sin k\sigma_{z}\otimes\tau_{y}.\label{eq:hbotk}
\end{alignat}

Since $J_{y}$ and $J_{d}$ are in general complex parameters of the
system, $U(k)$ is not unitary. In terms of the biorthogonal eigenbasis
of $U(k)$, Eq.~(\ref{eq:Uk}) can be equivalently written as

\begin{equation}
U(k)=\sum_{\ell=1,2}\sum_{\eta=\pm}e^{-{\rm i}\varepsilon_{\ell}^{\eta}(k)}|\varepsilon_{\ell}^{\eta}(k)\rangle\langle\overline{\varepsilon}_{\ell}^{\eta}(k)|,\label{eq:UkBio}
\end{equation}
where $|\varepsilon_{\ell}^{\eta}(k)\rangle$ ($\langle\overline{\varepsilon}_{\ell}^{\eta}(k)|$)
is the right (left) eigenvector of $U(k)$ with the quasienergy $\varepsilon_{\ell}^{\eta}(k)=\eta\varepsilon_{\ell}(k)\in\mathbb{C}$.
$\ell=1,2$ are the indices of the two quasienergy bands, whose real
parts satisfy ${\rm Re}[\varepsilon_{\ell}(k)]\in(0,\pi]$. The complex
dispersion $\{\varepsilon_{\ell}^{\eta}(k)\}$ thus contains four
Floquet bands, with two possible spectral gaps at quasienergies zero
and $\pi$. A topological phase transition may happen when a gap closes
at one of these quasienergies. This can be further captured by the vanishing
of one of the two gap functions $\Delta_0$ and $\Delta_\pi$, defined as

\begin{alignat}{1}
\Delta_{0} & \equiv\min_{k,\ell}\sqrt{[{\rm Re}\varepsilon_{\ell}(k)]^{2}+[{\rm Im}\varepsilon_{\ell}(k)]^{2}},\label{eq:D0}\\
\Delta_{\pi} & \equiv\min_{k,\ell}\sqrt{[|{\rm Re}\varepsilon_{\ell}(k)|-\pi]^{2}+[{\rm Im}\varepsilon_{\ell}(k)]^{2}}.\label{eq:DP}
\end{alignat}
In the next section, these functions will be utilized to obtain the boundaries between different Floquet topological phases of the non-Hermitian PQTLL model.

The topological invariants of the non-Hermitian Floquet phases
in our system are determined by the symmetries of $U(k)$. Following
the usual strategy in the study of one-dimensional (1D) Floquet systems \cite{AsbothSTF1,AsbothSTF2},
we rewrite $U(k)$ in a pair of symmetric time frames as

\begin{alignat}{1}
U_{1}(k)= & e^{-\frac{{\rm i}}{2}h_{\parallel}(k)}e^{-{\rm i}h_{\bot}(k)}e^{-\frac{{\rm i}}{2}h_{\parallel}(k)}=e^{-{\rm i}h_{1}(k)},\label{eq:U1k}\\
U_{2}(k)= & e^{-\frac{{\rm i}}{2}h_{\bot}(k)}e^{-{\rm i}h_{\parallel}(k)}e^{-\frac{{\rm i}}{2}h_{\bot}(k)}=e^{-{\rm i}h_{2}(k)}.\label{eq:U2k}
\end{alignat}
It is clear that $U_{1,2}(k)$ and $U(k)$ are related by similarity
transformations, which can be achieved by shifting the initial time
of the driving forward or backward over half a period. The Floquet
operators $U_{1,2}(k)$ thus share the same quasienergy dispersion
with $U(k)$, and they can be expressed in their corresponding biorthogonal
basis and as

\begin{equation}
U_{\alpha}(k)=\sum_{\ell=1,2}\sum_{\eta=\pm}e^{-{\rm i}\varepsilon_{\ell}^{\eta}(k)}|\varepsilon_{\alpha\ell}^{\eta}(k)\rangle\langle\overline{\varepsilon}_{\alpha\ell}^{\eta}(k)|,\label{eq:Uak}
\end{equation}
where $\alpha=1,2$ denote the two time frames. Moreover, the effective
Hamiltonians $h_{1,2}(k)$ in Eqs. (\ref{eq:U1k}) and (\ref{eq:U2k})
both possess the extended time-reversal symmetry ${\cal T}$, the extended particle hole symmetry ${\cal C}$, and the sublattice symmetry ${\cal S}$, i.e., 

\begin{equation}
{\cal T}={\rm i}\sigma_{y}\otimes\tau_{0},\qquad{\cal T}{\cal T}^{*}=-1,\qquad{\cal T}h_{\alpha}^{\top}(k){\cal T}^{-1}=h_{\alpha}(-k),
\end{equation}
\begin{equation}
{\cal C}=\sigma_{x}\otimes\tau_{y},\qquad{\cal C}{\cal C}^{*}=-1,\qquad{\cal C}h_{\alpha}^{\top}(k){\cal C}^{-1}=-h_{\alpha}(-k),
\end{equation}
\begin{equation}
{\cal S}=\sigma_{z}\otimes\tau_{y},\qquad{\cal S}^{2}=1,\qquad{\cal S}h_{\alpha}(k){\cal S}=-h_{\alpha}(k).
\end{equation}
According to the symmetry classification of Floquet systems \cite{STPTab,FTPTab}
and the periodic table of non-Hermitian topological phases \cite{Class1},
the non-Hermitian PQTLL model belongs to an extended CII symmetry
class with even-integer topological invariants. In the meantime, the
system also possesses the inversion symmetry ${\cal P}=\sigma_{x}\otimes\tau_{0}$
with ${\cal P}^{2}=1$, in the sense that ${\cal P}h_{\alpha}(k){\cal P}^{-1}=h_{\alpha}(-k)$
for $\alpha=1,2$. According to Ref.~\cite{Class1}, the coexistence of time-reversal
and inversion symmetries allows a system to be immune to the non-Hermitian
skin effect \cite{NHSkin}. The topological characterization and bulk-boundary
correspondence of our non-Hermitian PQTLL model can thus be treated
in a standard manner, as will be presented in the following sections.

\section{Topological invariants and phase diagrams\label{sec:WN}}
In this section, we introduce the topological invariants of our non-Hermitian
PQTLL model, and construct its topological phase diagrams for typical
situations.

Following the symmetry analysis in the last section and the topological
characterizations of Hermitian Floquet phases \cite{FSCL,SKR}, the Floquet
operator $U_{\alpha}(k)$ in the $\alpha$'s time frame possesses
a topological winding number $w_{\alpha}$, which can be defined as

\begin{equation}
w_{\alpha}=\int_{-\pi}^{\pi}\frac{dk}{4\pi}{\rm Tr}[{\cal S}{\cal Q}_{\alpha}(k){\rm i}\partial_{k}{\cal Q}_{\alpha}(k)],\label{eq:Wa}
\end{equation}
where $\alpha=1,2$, $k$ is the quasimomentum, ${\cal S}$ is the
sublattice symmetry operator, and the trace is taken over all the
internal degrees of freedom including spins and sublattices. ${\cal Q}_{\alpha}(k)$
is usually called the ${\cal Q}$-matrix~\cite{QMatrix}, which takes the form
of a biorthogonal projector

\begin{equation}
{\cal Q}_{\alpha}(k)=\sum_{\ell,\eta}\eta|\varepsilon_{\alpha\ell}^{\eta}(k)\rangle\langle\overline{\varepsilon}_{\alpha\ell}^{\eta}(k)|.\label{eq:Qak}
\end{equation}
Here $\ell=1,2$ are the indices of the two Floquet quasienergy bands,
whose real parts are positive. $\eta=\pm$ denote the signs of the
real parts of quasienergies. The right (left) eigenvectors $\{|\varepsilon_{\alpha\ell}^{\eta}(k)\rangle\}$
($\{|\overline{\varepsilon}_{\alpha\ell}^{\eta}(k)\rangle\}$) can
be obtained by expressing the Floquet operator in the $\alpha$'s time frame as $U_{\alpha}(k)=V_{\alpha}(k)\Lambda_{\alpha}(k)V_{\alpha}^{-1}(k)$,
where $\Lambda_{\alpha}(k)$ is diagonal and $\{|\varepsilon_{\alpha\ell}^{\eta}(k)\rangle\}$
($\{|\overline{\varepsilon}_{\alpha\ell}^{\eta}(k)\rangle\}$) are
given by the columns of $V_{\alpha}(k)$ ($[V_{\alpha}^{-1}(k)]^{\dagger}$)~\cite{QMatrix}.
Therefore, ${\cal Q}_{\alpha}(k)$ can be viewed as a flattened effective
Hamiltonian of $U_{\alpha}(k)$, whose eigenphases with positive and
negative real parts are set to zero and $\pi$, respectively.

With the help of $(w_{1},w_{2})$ in Eq.~(\ref{eq:Wa}) and the topological characterization of chiral symmetric Floquet systems \cite{AsbothSTF2}, we can construct
another pair of topological winding numbers $(w_{0},w_{\pi})$ as

\begin{equation}
w_{0}=\frac{w_{1}+w_{2}}{2},\qquad w_{\pi}=\frac{w_{1}-w_{2}}{2}.\label{eq:W0P}
\end{equation}
According to Ref.~\cite{FSCL}, these invariants would only take even-integer
values, and they provide a complete characterization for all
1D Hermitian Floquet topological phases in the CII symmetry class. Furthermore,
the requirement of two invariants reveals the difference between Floquet
and non-driven systems. Since the Floquet operator $U$ possesses
two quasienergy gaps at $\varepsilon=0$ and $\pi$, there could be two types
of degenerate edge modes at these quasienergies, whose numbers are
thus counted by two distinct topological invariants. In the following,
we will demonstrate that the topological invariants $(w_{0},w_{\pi})$
in Eq.~(\ref{eq:W0P}) could also characterize the Floquet phases
of the non-Hermitian PQTLL model, and they always take real and even-integer
values for a gapped topological phase.

\begin{figure}
	\centering
	\includegraphics[scale=0.5]{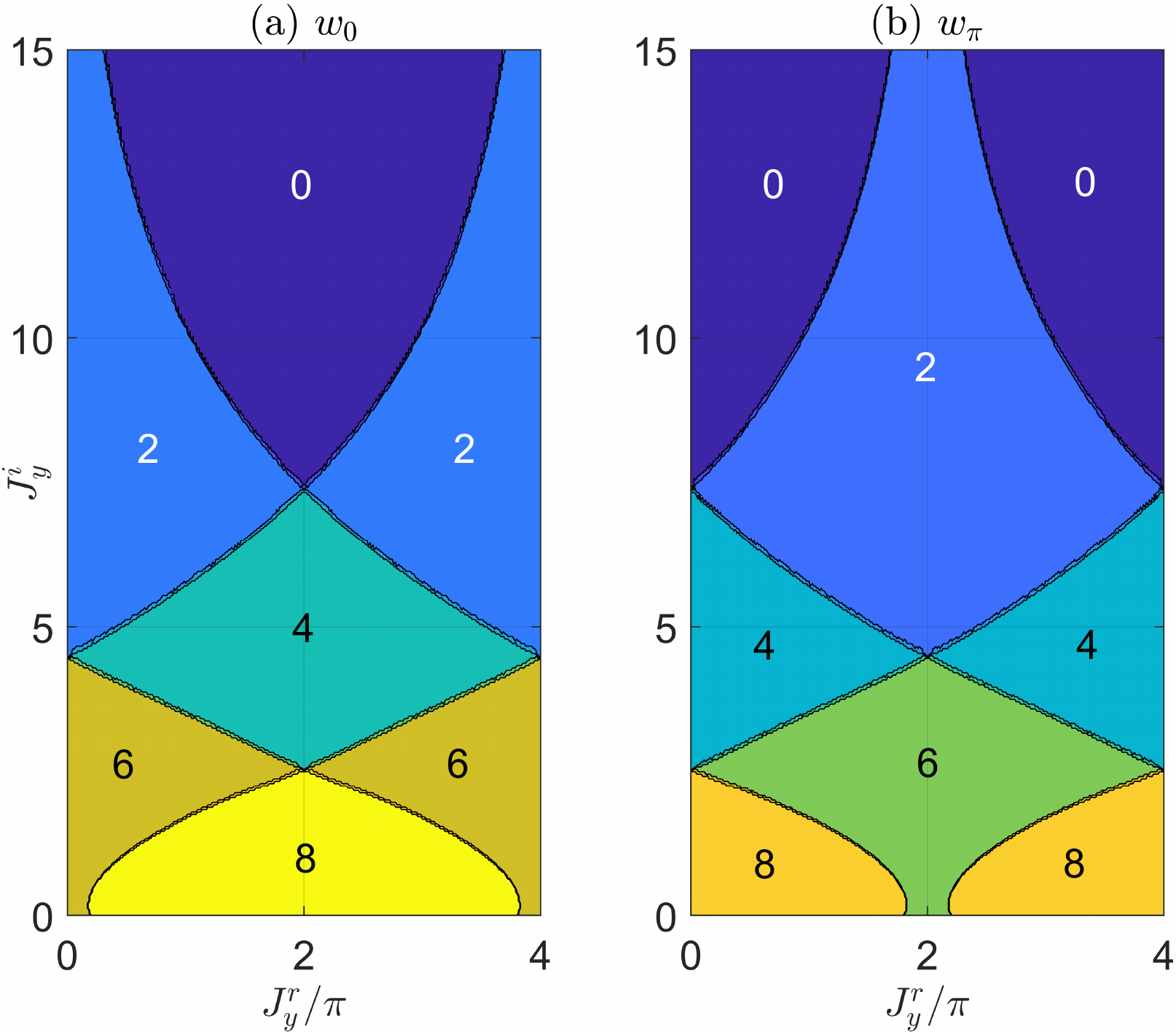}
	\caption{The topological winding numbers $w_{0}$ {[}in panel \textbf{(a)}{]} and $w_{\pi}$
		{[}in panel \textbf{(b)}{]} versus the real and imaginary parts of the vertical
		hopping amplitude $J_{y}^{r}$ and $J_{y}^{i}$. The other system
		parameters are chosen as $(J_{x},J_{d},V)=(0.5\pi,4\pi,0.1\pi)$.
		In both panels, each region with a uniform color corresponds to a
		Floquet topological phase of the non-Hermitian PQTLL model, with the
		values of winding numbers $(w_0,w_\pi)$ denoted explicitly therein. The lines separating
		different regions are the boundaries between different topological
		phases, which can be obtained numerically from the gap closing conditions
		$\Delta_{0}=0$ and $\Delta_{\pi}=0$ with the help of Eqs.~(\ref{eq:D0})
		and (\ref{eq:DP}). \label{fig:PhsDiag1}}
\end{figure}

\begin{figure}
	\centering
	\includegraphics[scale=0.5]{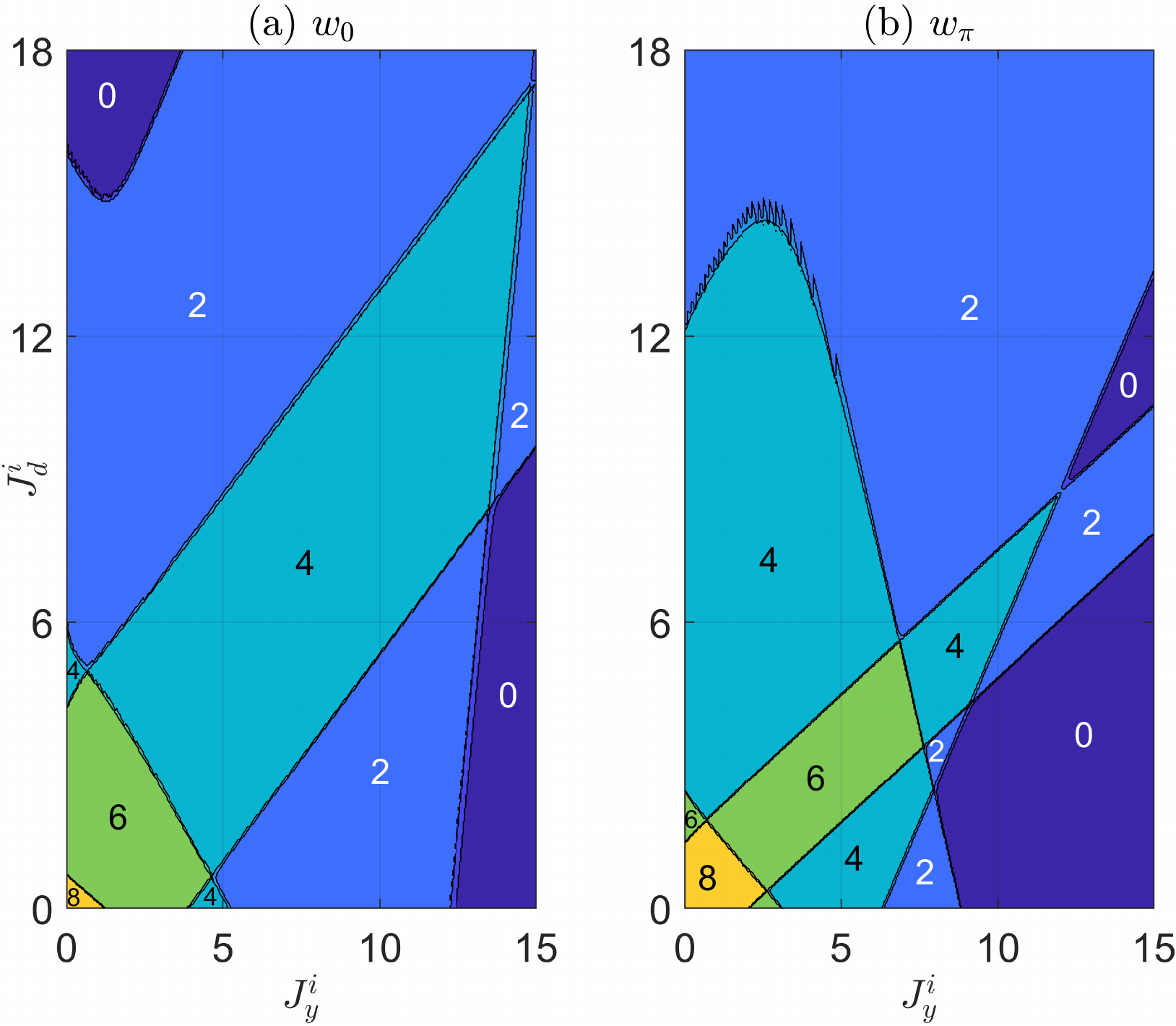}
	\caption{The topological winding numbers $w_{0}$ {[}in panel (a){]} and $w_{\pi}$
		{[}in panel (b){]} versus the imaginary parts of vertical and diagonal
		hopping amplitudes $J_{y}^{i}$ and $J_{d}^{i}$. The other system
		parameters are set as $(J_{x},J_{y}^{r},J_{d}^{r},V)=(0.5\pi,0.6\pi,4\pi,0.1\pi)$.
		In both panels, each region with a uniform color refers to a Floquet
		topological phase of the non-Hermitian PQTLL model, with the values
		of winding numbers $(w_0,w_\pi)$ shown explicitly in the figure. The lines separating
		different regions are the boundaries between different non-Hermitian
		Floquet topological phases, which can be obtained by setting $\Delta_{0}=0$
		and $\Delta_{\pi}=0$ in Eqs. (\ref{eq:D0}) and (\ref{eq:DP}). \label{fig:PhsDiag2}}
\end{figure}

By evaluating $(w_{0},w_{\pi})$ numerically with Eqs.~(\ref{eq:Wa})
and (\ref{eq:Qak}), we obtain the topological phase diagrams of
the non-Hermitian PQTLL model for two typical cases, as presented
in Figures \ref{fig:PhsDiag1} and \ref{fig:PhsDiag2}. In Figure \ref{fig:PhsDiag1},
we show the phase diagram of the system with respect to the real and
imaginary parts of the vertical hopping amplitude $J_{y}^{r}$ and
$J_{y}^{i}$. The other system parameters are all chosen to be real
and set as $(J_{x},J_{d},V)=(0.5\pi,4\pi,0.1\pi)$. From the phase
diagram, we see clearly that with the increase of the nonreciprocal
parameter $J_{y}^{i}$, a series of topological phase transitions
can be induced, with each of them being followed by the quantized
change of $w_{0}$ or $w_{\pi}$ by an integer multiple of two. The
resulting non-Hermitian Floquet topological phases could possess
large and even-integer topological invariants due to the interplay
between drivings and non-Hermitian effects. Moreover, phases carrying
larger topological winding numbers can be realized when the diagonal
hopping amplitude $J_{d}$ takes larger values. Therefore, the realization
of non-Hermitian PQTLL model could also provide us with a convenient
platform to explore non-Hermitian phases with large and even-integer
topological numbers.
In Figure \ref{fig:PhsDiag2}, we present the topological phase diagram of
the non-Hermitian PQTLL model versus the imaginary parts of the vertical
and diagonal hopping amplitudes $J_{y}^{i}$ and $J_{d}^{i}$. The
other system parameters are fixed at $(J_{x},J_{y}^{r},J_{d}^{r},V)=(0.5\pi,0.6\pi,4\pi,0.1\pi)$.
From the phase diagram, we again observe rich non-Hermitian Floquet
topological phases characterized by $(w_{0},w_{\pi})\in2\mathbb{Z}\times2\mathbb{Z}$,
and multiple topological phase transitions induced by the change of
the two non-Hermitian parameters. Furthermore, in certain regions
of the phase diagram (e.g., around $J_{y}^{i}=6$), we find phase
transitions accompanied by the increase of topological winding numbers
$(w_{0},w_{\pi})$ when the value of $J_{d}^{i}$ raises. The emergence
of such phases with stronger topological signatures in deeper non-Hermitian
regimes (here at larger $J_{d}^{i}$) is \emph{unique} to Floquet non-Hermitian
systems. In the meantime, it also suggests an approach to prepare
topological phases with large winding numbers under the collaboration
of drivings and nonreciprocity. 
In the following two sections, we will present the edge states and
bulk dynamics of the non-Hermitian PQTLL model, which would provide
us with more transparent signatures about its topological properties.

\section{Edge states and bulk-edge correspondence\label{sec:BBC}}
One of the key features for a gapped topological phase is the presence
of degenerate edge states under the OBC~\cite{BBC1}. In this section, we demonstrate
the existence of Floquet topological edge modes at zero- and $\pi$-quasienergies
in our non-Hermitian PQTLL model, and relate their numbers to the
bulk topological winding numbers $(w_{0},w_{\pi})$ in Eq.~(\ref{eq:W0P}).

The Floquet quasienergy spectrum of our system under the OBC is obtained
by solving the quasienergy eigenvalue equation $U|\psi\rangle=e^{-{\rm i}\varepsilon}|\psi\rangle$, with the Floquet operator $U$ given by Eq.~(\ref{eq:U}). With the
quasienergy $\varepsilon$, we can define the (point) gap functions
under the OBC as

\begin{equation}
\Delta_{0} \equiv\sqrt{({\rm Re}\varepsilon)^{2}+({\rm Im}\varepsilon)^{2}},\qquad\Delta_{\pi}  \equiv\sqrt{(|{\rm Re}\varepsilon|-\pi)^{2}+({\rm Im}\varepsilon)^{2}}.\label{eq:D0POBC}
\end{equation}
It is clear that $\Delta_{0}=0$ ($\Delta_{\pi}=0$) only when the
spectrum gap closes at the quasienergy $0$ ($\pi$). $(\Delta_{0},\Delta_{\pi})$
can thus be used to characterize the behaviors of the Floquet spectrum
around the quasienergies $\varepsilon=0$ and $\pi$.

\begin{figure}
	\centering
	\includegraphics[scale=0.5]{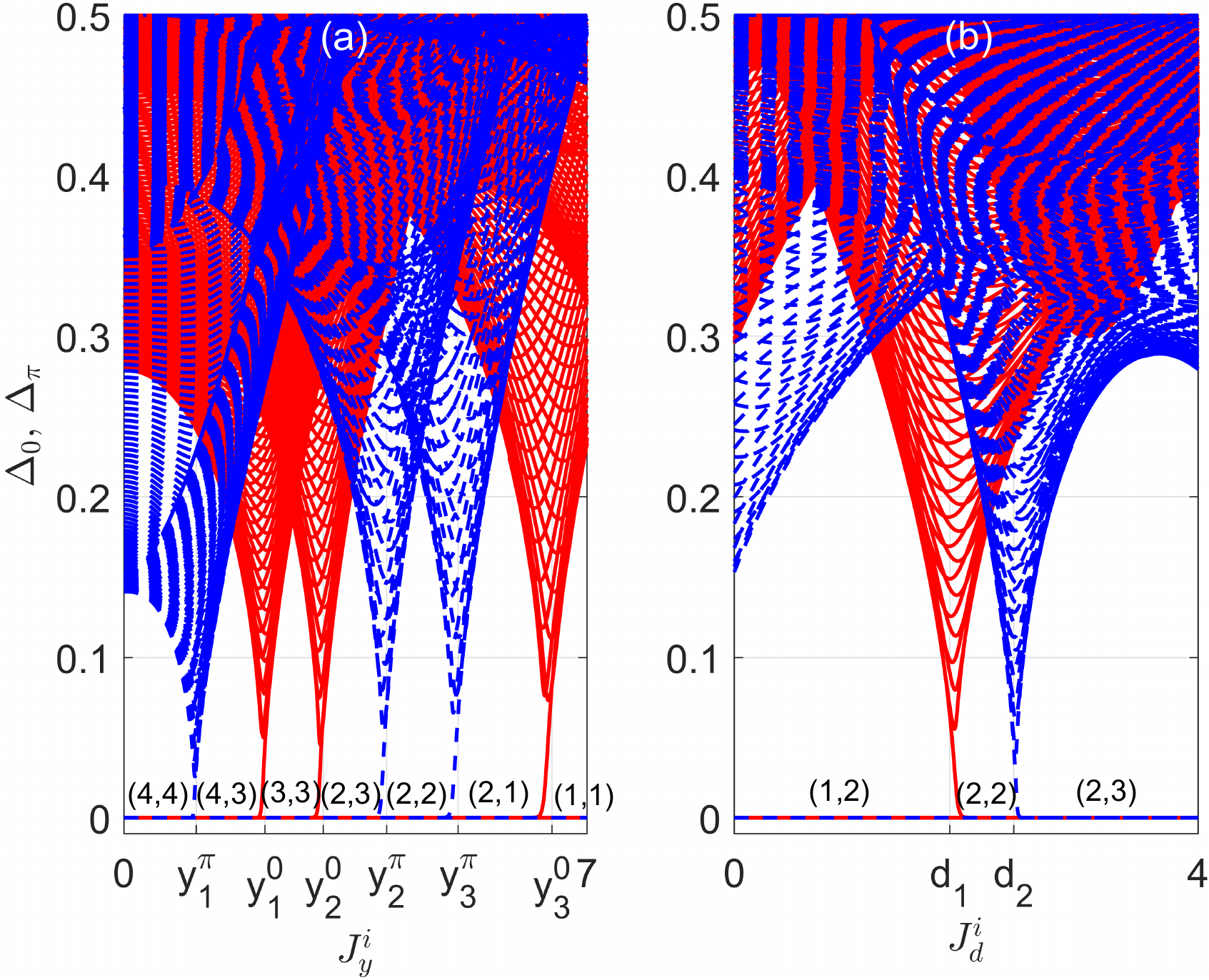}
	\caption{Gap functions $\Delta_{0}$ (red solid lines) and $\Delta_{\pi}$
		(blue dashed lines) versus the imaginary part of vertical and diagonal
		hopping amplitudes $J_{y}^{i}$ and $J_{d}^{i}$ in panels (a) and
		(b), respectively. The system parameters are $(J_{x},J_{y}^{r},J_{d},V)=(0.5\pi,1.5\pi,4\pi,0.1\pi)$
		for panel (a) and $(J_{x},J_{y},J_{d}^{r},V)=(0.5\pi,0.6\pi+6{\rm i},4\pi,0.1\pi)$
		for panel (b). The number of quartets of zero and $\pi$ edge modes
		$(n_{0},n_{\pi})$ are denoted explicitly near $\Delta_{0}=\Delta_{\pi}=0$
		in both panels, which are related to the winding numbers $(w_{0},w_{\pi})$
		through the relations in Eq.~(\ref{eq:BBC1}). The ticks along the
		horizontal axis denote the bulk gap closing points, whose numerical values are
		$(y_{1}^{\pi},y_{1}^{0},y_{2}^{0},y_{2}^{\pi},y_{3}^{\pi},y_{3}^{0})\approx(1.09,2.14,3.02,3.97,5.05,6.47)$
		in panel (a) and $(d_{1},d_{2})\approx(1.86,2.41)$ in panel (b).\label{fig:EOBC}}
\end{figure}

In Figures \ref{fig:EOBC}(a) and \ref{fig:EOBC}(b), we present the gap functions
$\Delta_{0}$ (red solid lines) and $\Delta_{\pi}$ (blue dashed lines)
of the non-Hermitian PQTLL model versus the imaginary parts of the
vertical and diagonal hopping amplitudes $J_{y}^{i}$ and $J_{d}^{i}$
for two typical sets of system parameters, respectively. In both panels,
we see clearly that with the increase of the nonreciprocal hopping
parameter $J_{y}^{i}$ or $J_{d}^{i}$, the system undergoes a series
of topological phase transitions. Each transition is accompanied by
the closing and reopening of a point gap at the quasienergy zero or
$\pi$, together with the increase or decrease of the number of Floquet
zero or $\pi$ edge modes by an integer multiple of four, as denoted
in the figure. Intriguingly, by enhancing the strength of nonreciprocity, we observe transitions from topological
phases with smaller winding numbers $(w_{0},w_{\pi})$ to larger ones
with more edge modes in Figure \ref{fig:EOBC}(b). This is unique to non-Hermitian Floquet
systems thanks to the interplay between drivings and non-Hermitian
effects. This observation also indicate the possibility of preparing
non-Hermitian Floquet topological phases with the help of nonreciprocity.

Furthermore, comparing the number of quartets of the zero ($\pi$)
edge modes $n_{0}$ ($n_{\pi}$) and the bulk winding number $w_{0}$ ($w_{\pi}$)
in each regime of the non-Hermitian Floquet topological phase, we
find the following bulk-edge correspondence relations

\begin{equation}
|w_{0}|=2n_{0},\qquad|w_{\pi}|=2n_{\pi}.\label{eq:BBC1}
\end{equation}
These relations hold so long as the symmetries that protecting the
non-Hermitian Floquet topological phases of the system are preserved.
Experimentally, Eq.~(\ref{eq:BBC1}) could also provide us with a window
to look into the topological invariants of non-Hermitian Floquet systems
in the CII symmetry class by imaging the edge modes. More generally,
in the symmetric time frame $\alpha$ ($=1,2$), we can directly define
a noncommutative winding number \cite{FSCL,QMatrix,QMatrix2,QMatrix3} under the OBC as

\begin{equation}
\breve{w}_{\alpha}\equiv\frac{1}{2N_{B}}{\rm Tr}_{B}({\cal S}{\cal Q}_{\alpha}[{\cal Q}_{\alpha},\hat{n}]),\label{eq:WaOBC}
\end{equation}
where ${\cal S}$ is again the sublattice symmetry operator, and $\hat{n}=\sum_{n=1}^{N}n|n\rangle\langle n|\sigma_{0}\otimes\tau_{0}$
is the unit-cell position operator of the ladder. The total number
of unit cells $N$ contains two parts, i.e., $N=N_{B}+2N_{E}$, where
$N_{B}$ and $N_{E}$ are the number of unit cells in the bulk ($n\in[N_{E}+1,N_{E}+N_{B}]$)
and edge ($n\in[1,N_{E}]\cup[N-N_{E}+1,N]$) regions of the system,
and the trace ${\rm Tr}_{B}(\cdot)$ is only taken over the bulk degrees
of freedom. Different from the previous study \cite{FSCL}, the ${\cal Q}$-matrix
for our non-Hermitian Floquet system in the $\alpha$'s time frame
and under the OBC is expressed in the biorthogonal basis as

\begin{equation}
{\cal Q}_{\alpha}\equiv\sum_{n,\eta}\eta|\varepsilon_{\alpha n}^{\eta}\rangle\langle\overline{\varepsilon}_{\alpha n}^{\eta}|,\label{eq:QaOBC}
\end{equation}
where $n=1,2,...,2N$, $\eta=\pm$, and $|\varepsilon_{\alpha n}^{\eta}\rangle$
is the right Floquet eigenvector satisfying $U_{\alpha}|\varepsilon_{\alpha n}^{\eta}\rangle=e^{-{\rm i}\varepsilon_{n}^{\eta}}|\varepsilon_{\alpha n}^{\eta}\rangle$.
The left eigenvectors can be obtained by writing $U_{\alpha}$ as
$U_{\alpha}=V_{\alpha}\Lambda_{\alpha}V_{\alpha}^{-1}$, where $\Lambda_{\alpha}$
is diagonal and $\{|\overline{\varepsilon}_{\alpha n}^{\eta}\rangle\}$
correspond to the columns of $(V_{\alpha}^{-1})^{\dagger}$ \cite{QMatrix}. Similar
to the ${\cal Q}$-matrix under the PBC, ${\cal Q}_{\alpha}$ here
can be viewed as an effective Hamiltonian of the Floquet operator
$U_{\alpha}$, whose eigenvalues are set to $1$ ($-1$) if the corresponding
quasienergies of $U_{\alpha}$ satisfying ${\rm Re}(\varepsilon_{n}^{\eta})>0$
[${\rm Re}(\varepsilon_{n}^{\eta})<0$]. With the help of $\breve{w}_{1}$
and $\breve{w}_{2}$ in Eq.~(\ref{eq:QaOBC}), we can construct another pair of topological
invariants \cite{FSCL}

\begin{equation}
\breve{w}_{0}=\frac{\breve{w}_{1}+\breve{w}_{2}}{2},\qquad\breve{w}_{\pi}=\frac{\breve{w}_{1}-\breve{w}_{2}}{2}.\label{eq:W0POBC}
\end{equation}
In a fixed time frame $\alpha$, previous studies have showed that
$\breve{w}_{\alpha}=w_{\alpha}$ \cite{FSCL}. Therefore, we find the following
bulk-edge correspondence relations for 1D non-Hermitian Floquet systems
in the extended CII symmetry class, i.e.,

\begin{equation}
(|w_{0}|,|w_{\pi}|)=(|\breve{w}_{0}|,|\breve{w}_{\pi}|)=(2n_{0},2n_{\pi}).\label{eq:BBC2}
\end{equation}
Since the second equality holds also under the OBC, it can be employed
to investigate the effect of impurity on non-Hermitian Floquet topological
phases, and characterize the non-Hermitian Floquet Anderson insulators
that may appear due to the interplay between drivings, non-Hermiticity
and disorder. These topics are beyond the scope of the current work,
and will be explored in the future. 

Despite edge states, the topological signatures of non-Hermitian Floquet
phases can also be extracted from bulk dynamics, as will be discussed
in the next section.

\section{Dynamical probe to the topological phases\label{sec:MCD}}
The mean chiral displacement (MCD) refers to the time-averaged chiral
displacement ${\cal S}\hat{n}$ of a wavepacket in a lattice, where
${\cal S}$ is the sublattice symmetry operator and $\hat{n}$ is
the position operator of the unit cell. The MCD was first introduced as
a dynamical probe to the winding numbers of 1D topological
insulators in the symmetry classes AIII and BDI \cite{MCD1}, and later extended
to Floquet systems \cite{SKR,MCD2}, two-dimensional systems \cite{MCD3}, many-body
systems \cite{MCD4}, systems in other symmetry classes \cite{FSCL}, and recently
also to non-Hermitian systems \cite{ZhouNHFTP2,ZhouNHFTP3,ZhouNHFTP5}. In the meantime, the MCD has
also been measured experimentally in photonic \cite{MCD1,MCD5} and cold atom
\cite{MCD6,MCD7} setups. In this section, we further generalize the MCD to non-Hermitian
Floquet systems in the CII symmetry class, and employ it to dynamically
characterize the topological phases found in the non-Hermitian PQTLL
model. 

For a non-Hermitian Floquet system with sublattice symmetry ${\cal S}$,
we define the MCD $C_{\alpha}$ as the stroboscopic long-time average
of the chiral displacement operator ${\cal S}\hat{n}$ in a given
symmetric time frame $\alpha$ ($=1,2$), i.e.,

\begin{equation}
C_{\alpha}=\lim_{M\rightarrow\infty}\frac{1}{M}\sum_{m=1}^{M}\langle\overline{\psi}(m)|{\cal S}\hat{n}|\psi(m)\rangle,\label{eq:MCD0}
\end{equation}
where $m$ counts the number of driving periods, which has been set
to $1$ following our choice of units. $|\psi(m)\rangle$ and $|\overline{\psi}(m)\rangle$
are the initial states evolved over $m$'s driving periods in the
right and left Hilbert spaces, respectively. The $C_{\alpha}$ defined
in this way is generally a complex number for finite $M$ due to the
implemented biorthogonal average. However, we will show that the imaginary
part of $C_{\alpha}$ tend to vanish in the long-time limit $M\rightarrow\infty$.

Taking the Fourier transform from the position to momentum representation,
and choosing the initial state to uniformly fill the non-Hermitian
quasienergy band $(\ell,\eta)$ ($\ell=1,2$, $\eta=\pm$), we find
the following form of MCD

\begin{alignat}{1}
C_{\alpha\ell}^{\eta}& =  \lim_{M\rightarrow\infty}\frac{1}{M}\sum_{m=1}^{M}\int_{-\pi}^{\pi}\frac{dk}{2\pi}c_{\alpha\ell}^{\eta}(k),\label{eq:Cal0}\\
c_{\alpha\ell}^{\eta}(k)& =  \frac{\langle\overline{\varepsilon}_{\alpha\ell}^{\eta}(k)|\overline{U}_{\alpha}^{m\dagger}(k){\cal S}{\rm i}\partial_{k}U_{\alpha}^{m}(k)|\varepsilon_{\alpha\ell}^{\eta}(k)\rangle}{\langle\overline{\varepsilon}_{\alpha\ell}^{\eta}(k)|\overline{U}_{\alpha}^{m\dagger}(k)U_{\alpha}^{m}(k)|\varepsilon_{\alpha\ell}^{\eta}(k)\rangle}.\label{eq:cal0}
\end{alignat}
Here $|\varepsilon_{\alpha\ell}^{\eta}(k)\rangle$ ($\langle\overline{\varepsilon}_{\alpha\ell}^{\eta}(k)|$)
is the right (left) quasienergy eigenvector, and the corresponding
Floquet operators can be expressed in the biorthogonal basis as

\begin{alignat}{1}
U_{\alpha}(k)= & \sum_{\ell,\eta}e^{-{\rm i}\varepsilon_{\ell}^{\eta}(k)}|\varepsilon_{\alpha\ell}^{\eta}(k)\rangle\langle\overline{\varepsilon}_{\alpha\ell}^{\eta}(k)|,\label{eq:Ua}\\
\overline{U}_{\alpha}^{\dagger}(k)= & \sum_{\ell,\eta}e^{+{\rm i}\varepsilon_{\ell}^{\eta*}(k)}|\varepsilon_{\alpha\ell}^{\eta}(k)\rangle\langle\overline{\varepsilon}_{\alpha\ell}^{\eta}(k)|.\label{eq:Ua-}
\end{alignat}
Note that in Eq.~(\ref{eq:Cal0}), a normalization factor has been
added to cancel the changing norm of the state during the nonunitary
evolution. Inserting the identity in biorthogonal basis $\mathbb{I}=\sum_{\ell,\eta}|\varepsilon_{\alpha\ell}^{\eta}(k)\rangle\langle\overline{\varepsilon}_{\alpha\ell}^{\eta}(k)|$,
and using the orthonormality between left and right eigenvectors $\langle\overline{\varepsilon}_{\alpha\ell}^{\eta}(k)|\varepsilon_{\alpha\ell'}^{\eta'}(k)\rangle=\delta_{\ell\ell'}\delta_{\eta\eta'}$,
the denominator of $c_{\alpha\ell}^{\eta}(k)$ in Eq.~(\ref{eq:cal0})
can be simplified as

\begin{equation}
\langle\overline{\varepsilon}_{\alpha\ell}^{\eta}(k)|\overline{U}_{\alpha}^{m\dagger}(k)U_{\alpha}^{m}(k)|\varepsilon_{\alpha\ell}^{\eta}(k)\rangle=e^{2{\rm Im}[\varepsilon_{\ell}^{\eta}(k)]m},\label{eq:Deno1}
\end{equation}
where ${\rm Im}[\varepsilon_{\ell}^{\eta}(k)]$ yields the
imaginary part of the quasienergy $\varepsilon_{\ell}^{\eta}(k)$.
Similarly, the numerator of $c_{\alpha\ell}^{\eta}(k)$ can be expressed
as

\begin{equation}
\langle\overline{\varepsilon}_{\alpha\ell}^{\eta}(k)|\overline{U}_{\alpha}^{m\dagger}(k){\cal S}{\rm i}\partial_{k}U_{\alpha}^{m}(k)|\varepsilon_{\alpha\ell}^{\eta}(k)\rangle=e^{2{\rm Im}[\varepsilon_{\ell}^{\eta}(k)]m}\langle\overline{\varepsilon}_{\alpha\ell}^{\eta}(k)|{\cal S}|{\rm i}\partial_{k}\varepsilon_{\alpha\ell}^{\eta}(k)\rangle-e^{{\rm i}2{\rm Re}[\varepsilon_{\ell}^{\eta}(k)]m}\langle\overline{\varepsilon}_{\alpha\ell}^{-\eta}(k)|{\cal S}|{\rm i}\partial_{k}\varepsilon_{\alpha\ell}^{-\eta}(k)\rangle,\label{eq:Nume1}
\end{equation}
where we have also used the fact ${\cal S}|\varepsilon_{\alpha\ell}^{\eta}(k)\rangle\propto|\varepsilon_{\alpha\ell}^{-\eta}(k)\rangle$
to arrive at the second term on the right hand side of Eq.~(\ref{eq:Nume1}).
Plugging Eqs.~(\ref{eq:Deno1}) and (\ref{eq:Nume1}) into Eq.~(\ref{eq:cal0}),
we find the integrand $c_{\alpha\ell}^{\eta}(k)$ to be

\begin{equation}
c_{\alpha\ell}^{\eta}(k)=\langle\overline{\varepsilon}_{\alpha\ell}^{\eta}(k)|{\cal S}|{\rm i}\partial_{k}\varepsilon_{\alpha\ell}^{\eta}(k)\rangle-\langle\overline{\varepsilon}_{\alpha\ell}^{-\eta}(k)|{\cal S}|{\rm i}\partial_{k}\varepsilon_{\alpha\ell}^{-\eta}(k)\rangle e^{{\rm i}2\varepsilon_{\ell}^{\eta}(k)m}.\label{eq:cal1}
\end{equation}
The first term on the right hand side of Eq.~(\ref{eq:cal1})
will be related to the winding number of the system in the $\alpha$'s
time frame. If ${\rm Im}[\varepsilon_{\ell}^{\eta}(k)]>0$, the second
term on right hand side of Eq.~(\ref{eq:cal1}) will vanish in general after taking
the sum over $m$ and the limit $M\rightarrow\infty$, as imposed
in Eq.~(\ref{eq:Cal0}). However, when ${\rm Im}[\varepsilon_{\ell}^{\eta}(k)]<0$,
the factor $e^{{\rm i}2\varepsilon_{\ell}^{\eta}(k)m}$ will contribute an exponentially growing factor to $c_{\alpha\ell}^{\eta}(k)$ after the summation over $m$,
making it diverge in the limit $M\rightarrow\infty$.

To remove the divergence, we introduce another pair of Floquet propagators
for the right and left initial states with ${\rm Im}[\varepsilon_{\ell}^{\eta}(k)]<0$,
which are given by

\begin{alignat}{1}
\acute{U}_{\alpha}(k)= & \sum_{\ell,\eta}e^{+{\rm i}\varepsilon_{\ell}^{\eta}(k)}|\varepsilon_{\alpha\ell}^{\eta}(k)\rangle\langle\overline{\varepsilon}_{\alpha\ell}^{\eta}(k)|=U_{\alpha}^{-1}(k),\label{eq:Uar}\\
\grave{U}_{\alpha}^{\dagger}(k)= & \sum_{\ell,\eta}e^{-{\rm i}\varepsilon_{\ell}^{\eta*}(k)}|\varepsilon_{\alpha\ell}^{\eta}(k)\rangle\langle\overline{\varepsilon}_{\alpha\ell}^{\eta}(k)|=[\overline{U}_{\alpha}^{\dagger}(k)]^{-1}.\label{eq:Ual}
\end{alignat}
Comapring with Eq. (\ref{eq:Ua}), It is clear that $\acute{U}_{\alpha}(k)$
is just the inverse of Floquet operator $U_{\alpha}(k)$, which
can be obtained by simply reversing the signs of all the system parameters
in our model. With these considerations, we modify the definition
of $C_{\alpha\ell}^{\eta}$ in Eq. (\ref{eq:Cal0}) to

\begin{equation}
C_{\alpha\ell}^{\eta}=\lim_{M\rightarrow\infty}\frac{1}{M}\sum_{m=1}^{M}\int_{-\pi}^{\pi}\frac{dk}{2\pi}\cdot\begin{cases}
c_{\alpha\ell}^{\eta}(k) & {\rm Im}[\varepsilon_{\ell}^{\eta}(k)]>0\\
\check{c}_{\alpha\ell}^{\eta}(k) & {\rm Im}[\varepsilon_{\ell}^{\eta}(k)]<0
\end{cases},\label{eq:Cal1}
\end{equation}
where $c_{\alpha\ell}^{\eta}(k)$ is given by Eq.~(\ref{eq:cal0}),
and $\check{c}_{\alpha\ell}^{\eta}(k)$ takes the form

\begin{equation}
\check{c}_{\alpha\ell}^{\eta}(k)=\frac{\langle\overline{\varepsilon}_{\alpha\ell}^{\eta}(k)|\grave{U}_{\alpha}^{m\dagger}(k){\cal S}{\rm i}\partial_{k}\acute{U}_{\alpha}^{m}(k)|\varepsilon_{\alpha\ell}^{\eta}(k)\rangle}{\langle\overline{\varepsilon}_{\alpha\ell}^{\eta}(k)|\grave{U}_{\alpha}^{m\dagger}(k)\acute{U}_{\alpha}^{m}(k)|\varepsilon_{\alpha\ell}^{\eta}(k)\rangle}.\label{eq:cal2}
\end{equation}
Following the steps in the derivations of Eqs. (\ref{eq:Deno1}) and
(\ref{eq:Nume1}), we find the denominator and numerator of $\check{c}_{\alpha\ell}^{\eta}(k)$
to be

\begin{equation}
\langle\tilde{\varepsilon}_{\alpha\ell}^{\eta}(k)|\grave{U}_{\alpha}^{m\dagger}(k)\acute{U}_{\alpha}^{m}(k)|\varepsilon_{\alpha\ell}^{\eta}(k)\rangle=e^{-2{\rm Im}[\varepsilon_{\ell}^{\eta}(k)]m},\label{eq:Deno2}
\end{equation}
\begin{equation}
\langle\overline{\varepsilon}_{\alpha\ell}^{\eta}(k)|\grave{U}_{\alpha}^{m\dagger}(k){\cal S}{\rm i}\partial_{k}\acute{U}_{\alpha}^{m}(k)|\varepsilon_{\alpha\ell}^{\eta}(k)\rangle=e^{-2{\rm Im}[\varepsilon_{\ell}^{\eta}(k)]m}\langle\overline{\varepsilon}_{\alpha\ell}^{\eta}(k)|{\cal S}|{\rm i}\partial_{k}\varepsilon_{\alpha\ell}^{\eta}(k)\rangle-e^{-{\rm i}2{\rm Re}[\varepsilon_{\ell}^{\eta}(k)]m}\langle\overline{\varepsilon}_{\alpha\ell}^{-\eta}(k)|{\cal S}|{\rm i}\partial_{k}\varepsilon_{\alpha\ell}^{-\eta}(k)\rangle.\label{eq:Nume2}
\end{equation}
Plugging them into Eq.~(\ref{eq:cal2}), we further obtain

\begin{equation}
\check{c}_{\alpha\ell}^{\eta}(k)=\langle\overline{\varepsilon}_{\alpha\ell}^{\eta}(k)|{\cal S}|{\rm i}\partial_{k}\varepsilon_{\alpha\ell}^{\eta}(k)\rangle-\langle\overline{\varepsilon}_{\alpha\ell}^{-\eta}(k)|{\cal S}|{\rm i}\partial_{k}\varepsilon_{\alpha\ell}^{-\eta}(k)\rangle e^{-{\rm i}2\varepsilon_{\ell}^{\eta}(k)m}.\label{eq:cal3}
\end{equation}
It is clear that under the condition ${\rm Im}[\varepsilon_{\ell}^{\eta}(k)]<0$,
the second term on the RHS of Eq.~(\ref{eq:cal3}) will in general
vanish under the summation and long-time average $\lim_{M\rightarrow\infty}\frac{1}{M}\sum_{m}$,
as imposed in Eq.~(\ref{eq:Cal1}).

Next, we extend the initial state of our system to an incoherent summation
of all uniformly filled Floquet bands $(\ell,\eta)$, which is equivalent
to an ``infinite-temperature'' state at each quasimomentum $k$.
In the lattice representation, such an initial state corresponds to
the uniform population of all the four sublattices in the central
unit cell of the ladder, which is relatively easy to prepare. For
such an initial state, the MCD becomes $C_{\alpha}=\sum_{\ell,\eta}C_{\alpha\ell}^{\eta}$.
With the help of Eqs. (\ref{eq:cal1}), (\ref{eq:Cal1}) and (\ref{eq:cal3}),
it can be written more compactly as

\begin{equation}
C_{\alpha}= \sum_{\ell,\eta}\int_{-\pi}^{\pi}\frac{dk}{2\pi}{\cal A}_{\alpha\ell}^{\eta}(k)\left[1-\lim_{M\rightarrow\infty}\frac{1}{M}\frac{1-e^{{\rm i}2s\varepsilon_{\ell}^{\eta}(k)M}}{e^{-{\rm i}2s\varepsilon_{\ell}^{\eta}(k)}-1}\right],\label{eq:Ca}
\end{equation}
where ${\cal A}_{\alpha\ell}^{\eta}(k)\equiv\langle\overline{\varepsilon}_{\alpha\ell}^{\eta}(k)|{\cal S}|{\rm i}\partial_{k}\varepsilon_{\alpha\ell}^{\eta}(k)\rangle$,
and $s\equiv{\rm sgn}\{{\rm Im}[\varepsilon_{\ell}^{\eta}(k)]\}$
refers to the sign of ${\rm Im}[\varepsilon_{\ell}^{\eta}(k)]$. It
is not hard to see that the second term on the right hand side of Eq.~(\ref{eq:Ca}) will
tend to vanish in the long-time limit $M\rightarrow\infty$, so long
as $\varepsilon_{\ell}^{\eta}(k)=\pm\pi/2,\pm\pi$ have solutions only
at a discrete set of $k$-points in the first Brillouin zone, which is the case for our system.

Finally, the relation between $C_{\alpha}$ and the winding number
$w_{\alpha}$ in the $\alpha$'s time frame can be established as
follows. Inserting the expression of projector ${\cal Q}_{\alpha}(k)$
in Eq.~(\ref{eq:Qak}) into the definition of $w_{\alpha}$ in Eq.~(\ref{eq:Wa}), and taking the trace in the biorthogonal basis, we find

\begin{equation}
w_{\alpha}=\int_{-\pi}^{\pi}\frac{dk}{4\pi}\sum_{\ell\ell',\eta\eta'}\eta\eta'\langle\overline{\varepsilon}_{\alpha\ell}^{\eta}(k)|{\rm i}\partial_{k}\left[|\varepsilon_{\alpha\ell'}^{\eta'}(k)\rangle\langle\overline{\varepsilon}_{\alpha\ell'}^{\eta'}(k)|\right]{\cal S}|\varepsilon_{\alpha\ell}^{\eta}(k)\rangle.\label{eq:Wa0}
\end{equation}
Using again the orthonormality between left and right eigenvectors
and the fact ${\cal S}|\varepsilon_{\alpha\ell}^{\eta}(k)\rangle\propto|\varepsilon_{\alpha\ell}^{-\eta}(k)\rangle$,
the expression for $w_{\alpha}$ can be simplified to

\begin{equation}
w_{\alpha}=\sum_{\ell,\eta}\int_{-\pi}^{\pi}\frac{dk}{2\pi{\rm i}}\langle\overline{\varepsilon}_{\alpha\ell}^{\eta}(k)|{\cal S}|\partial_{k}\varepsilon_{\alpha\ell}^{\eta}(k)\rangle.\label{eq:Wa1}
\end{equation}
Comparing Eq.~(\ref{eq:Wa1}) with Eq.~(\ref{eq:Ca}), we find the
relation between the long-time averaged MCD $C_{\alpha}$ and winding
number $w_{\alpha}$ as

\begin{equation}
w_{\alpha}=-C_{\alpha},\qquad\alpha=1,2.\label{eq:WaCa}
\end{equation}
Furthermore, with the help of the relations between $(w_{1},w_{2})$
and the topological invariants $(w_{0},w_{\pi})$ in Eq.~(\ref{eq:W0P}),
we arrive at the relations between the MCDs and the topological winding
numbers of 1D non-Hermitian Floquet systems in the CII symmetry
class, i.e.,

\begin{equation}
w_{0}=C_{0}\equiv-\frac{C_{1}+C_{2}}{2},\qquad w_{\pi}=C_{\pi}\equiv-\frac{C_{1}-C_{2}}{2}.\label{eq:W0PMCD}
\end{equation}
These relations establish a connection between the topology
and dynamics of the non-Hermitian Floquet states in the system. They also provide
us with a powerful way to probe the non-Hermitian Floquet topological
phases in the CII symmetry class by measuring the MCDs experimentally
in a pair of symmetric time frames.
\begin{figure}
	\centering
	\includegraphics[scale=0.5]{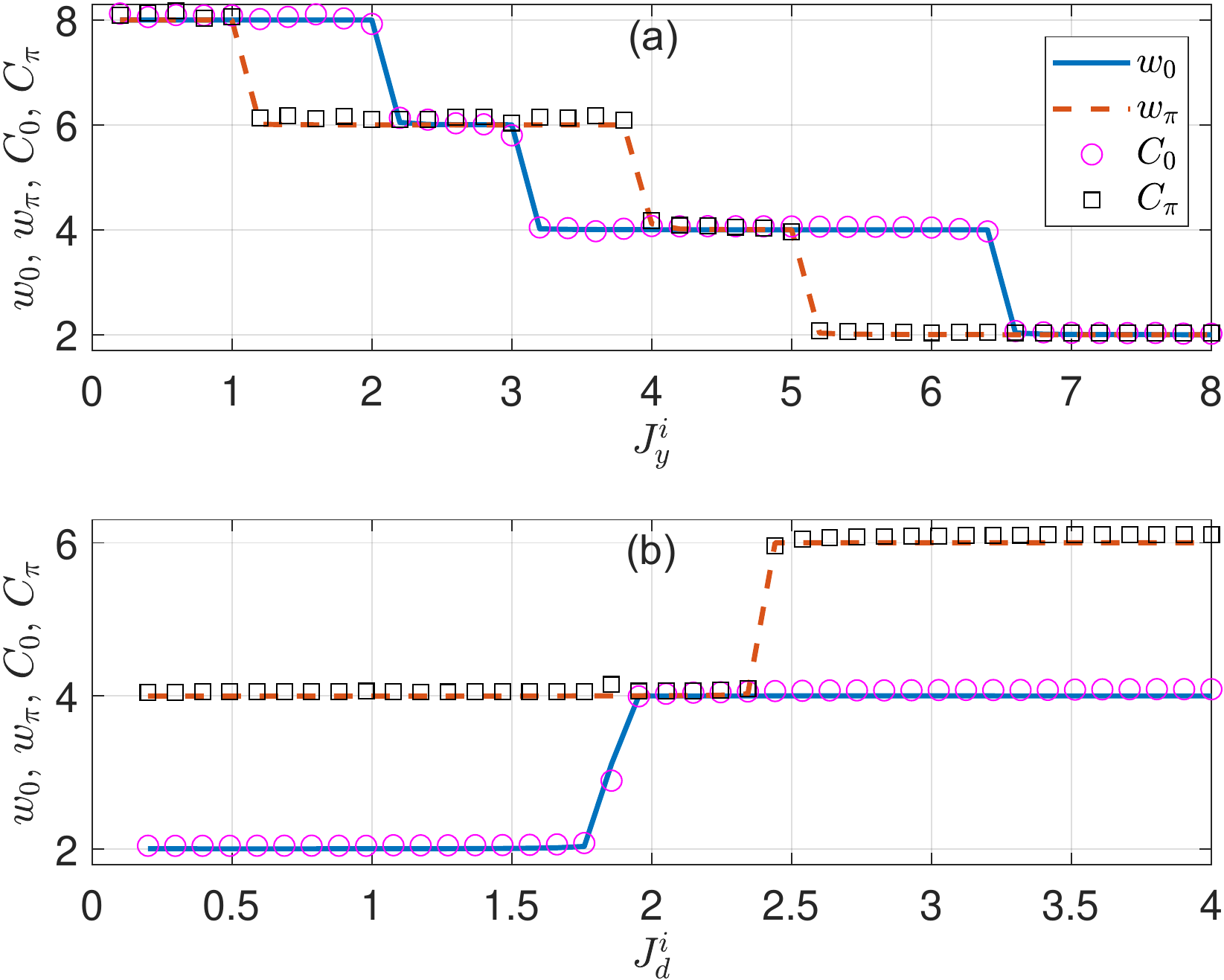}
	\caption{The topological winding numbers $w_{0}$ (blue solid lines), $w_{\pi}$
		(red dashed lines), MCDs $C_{0}=-\frac{C_{1}+C_{2}}{2}$ (magenta circles)
		and $C_{\pi}=-\frac{C_{1}-C_{2}}{2}$ (black squares) versus the imaginary
		parts of vertical and diagonal hopping amplitudes $J_{y}^{i}$ and
		$J_{d}^{i}$ of the non-Hermitian PQTLL model in panels (a) and (b),
		respectively. The other system parameters are chosen as $(J_{x},J_{y}^{r},J_{d},V)=(0.5\pi,1.5\pi,4\pi,0.1\pi)$
		for panel (a) and $(J_{x},J_{y},J_{d}^{r},V)=(0.5\pi,0.6\pi+6{\rm i},4\pi,0.1\pi)$
		for panel (b). The MCDs are averaged over $M=20$ driving periods
		for the results in both panels.\label{fig:MCD}}
\end{figure}
For completeness, we demonstrate the relations in Eq.~(\ref{eq:W0PMCD})
by numerically simulating the dynamics. The results for two typical
cases are represented in Figures \ref{fig:MCD}(a) and \ref{fig:MCD}(b). In both
panels, the time average is taken over $M=20$ driving periods, which
is well within reach in current experiments. It is clear that the
MCDs and topological winding numbers are consistent for all the non-Hermitian
Floquet topological phases considered in the figures, and the small
deviations are mainly originated from the finite-time effect. Furthermore,
a quantized jump of the MCD is observed every time when the system
passes through a topological phase transition point. Experimentally,
the MCDs have been measured in both the cold atom \cite{MCD6,MCD7} and photonic
systems \cite{MCD1,MCD5}, in which non-Hermiticity and driving fields can also
be implemented~\cite{NHRev1}. Furthermore, the MCDs may also be detected
directly in momentum space with the help of a recently proposed setup
certaining the nitrogen-vacancy-center in diamond \cite{NVCent1}. Putting
together, we conclude that the MCD can indeed be employed as a dynamical
probe to the topological phases and phase transitions in our non-Hermitian
PQTLL model, and also in other 1D non-Hermitian Floquet systems in
the extended CII symmetry class.

\section{Conclusions\label{sec:Summary}}
In this work, we introduced a periodically quenched two-leg ladder
model subjecting to nonreciprocal inter-leg hoppings. The system belongs
to an extended CII symmetry class in the non-Hermitian periodic table~\cite{Class1}, which is further characterized by a pair of even-integer topological
winding numbers $(w_{0},w_{\pi})\in2\mathbb{Z}\times2\mathbb{Z}$
due to the existence of time-periodic drivings. We established the
topological phase diagrams of the model, and observed rich non-Hermitian
Floquet topological phases with large winding numbers. Specially,
Floquet phases carrying larger topological invariants can emerge in
stronger non-Hermitian regimes thanks to the collaboration between
drivings and non-Hermiticity. Under the open boundary condition,
Floquet topological edge modes with zero and $\pi$ quasienergies
appear as fourfold degenerate quartets around the boundaries, whose exact numbers are
determined by the bulk topological invariants $(w_{0},w_{\pi})$.
Besides the bulk-edge correspondence, we introduced the generalized
mean chiral displacement as another probe to the topological features
of our system dynamically, and showed that the MCDs in long-time limit
yield the topological invariants of one-dimensional non-Hermitian
Floquet systems in the CII symmetry class. Our work not only uncovers
a new type of topological phase originated from the interplay between
drivings and non-Hermitian effects, but also paves the way for the dynamical characterization  of non-Hermitian Floquet
topological matter. In future work, it would be interesting to extend
our findings to other symmetry classes, higher spatial dimensions
and superconducting systems. Furthermore, intriguing non-Hermitian
Floquet phases and phenomena are expected to appear under the effects
of disorder and many-body interactions, which certainly deserve thorough
explorations.

\vspace{6pt} 





	





\begin{thebibliography}{99}
	\bibitem{NHRev1}
	Ashida, Y.; Gong, Z.; and Ueda, M. Non-Hermitian Physics. {\em arXiv:2006.01837} {\bf 2020}.
	
	\bibitem{NHRev2}
	Yoshida, T.; Peters, R.; Kawakami, N.; and Hatsugai, Y. Exceptional band touching for strongly correlated systems in equilibrium. {\em arXiv:2002.11265}  {\bf 2020}.
	
	\bibitem{NHRev3}
	Bergholtz, E.J.; Budich, J.C.; and Kunst, F.K. Exceptional Topology of Non-Hermitian Systems. {\em arXiv:1912.10048}  {\bf 2020}.
	
	\bibitem{NHRev4} Ghatak, A.; and Das, T. New topological invariants in non-Hermitian
	systems. {\em J. Phys.: Condens. Matter} {\bf 2019}, {\em 31}, 263001.
	
	\bibitem{NHRev5} Alvarez, V.M.M.; Vargas, J.E.B.; Berdakin, M.; and Foa Torres, L.E.F. Topological states of non-Hermitian systems. {\em Eur. Phys. J. Special Topics} {\bf 2018}, {\em 227}, 1295.
	
	\bibitem{NHRev6} El-Ganainy, R.; Makris, K.G.; Khajavikhan, M.; Musslimani, Z.H.; Rotter, S.; and Christodoulides, D.N. Non-Hermitian physics and PT symmetry. {\em Nat. Phys.} {\bf 2018}, {\em 14}, 11-19.
	
	\bibitem{NHRev7} Mostafazadeh, A.; and Batal, A. Physical aspects of pseudo-Hermitian and $PT$-symmetric quantum mechanics. {\em J. Phys. A: Math. Gen.} {\bf 2004}, {\em 37}, 11645-11679.
	
	\bibitem{Class1} Kawabata, K.; Shiozaki, K.; Ueda, M.; and Sato, M. Symmetry and Topology in Non-Hermitian Physics. {\em Phys. Rev. X} {\bf 2019}, {\em 9}, 041015.
	
	\bibitem{Class2} Gong, Z.; Ashida, Y.; Kawabata, K.; Takasan, K.; Higashikawa, S.; and Ueda, M.Topological Phases of Non-Hermitian Systems. {\em Phys. Rev. X} {\bf 2018}, {\em 8}, 031079.
	
	\bibitem{Class3} Zhou, H.; and Lee, J.Y. Periodic table for topological bands with non-Hermitian symmetries. {\em Phys. Rev. B} {\bf 2019}, {\em 99}, 235112.
	
	\bibitem{Class4} Shen, H.; Zhen, B.; and Fu, L. Topological Band Theory for Non-Hermitian Hamiltonians. {\em Phys. Rev. Lett.} {\bf 2018}, {\em 120}, 146402.
	
	\bibitem{Class5} Liu, C.-H.; and Chen, S. Topological classification of defects in non-Hermitian systems. {\em Phys. Rev. B} {\bf 2019}, {\em 100}, 144106.
	
	\bibitem{Class6} Lieu, S.; McGinley, M.; and Cooper, N.R. Tenfold Way for Quadratic Lindbladians. {\em Phys. Rev. Lett.} {\bf 2020}, {\em 124}, 040401.
	
	\bibitem{Class7} Wojcik, C.C.; Sun, X.-Q.; Bzdu${\check {\rm s}}$ek, T.; and Fan, S. Homotopy characterization of non-Hermitian Hamiltonians. {\em Phys. Rev. B} {\bf 2020}, {\em 101}, 205417.
	
	\bibitem{Dym1} Lee, J.Y.; Ahn, J.; Zhou, H.; and Vishwanath, A. Topological Correspondence between Hermitian and Non-Hermitian Systems: Anomalous Dynamics. {\em Phys. Rev. Lett.} {\bf 2019}, {\em 123}, 206404.
	
	\bibitem{Dym2} Zhou, L.; Wang, Q.-h.; Wang, H.; and Gong, J. Dynamical quantum phase transitions in non-Hermitian lattices. {\em Phys. Rev. A} {\bf 2018}, {\em 98}, 022129.
	
	\bibitem{Dym3} Zhu, B.; Ke, Y.; Zhong, H.; and Lee, C. Dynamic winding number for exploring band topology. {\em Phys. Rev. Research} {\bf 2020}, {\em 2}, 023043.
	
	\bibitem{Dym4} Doppler, J.; Mailybaev, A.A.; B\"ohm, J.; Kuhl, U.; Girschik, A.; Libisch, F.; and Milburn, T.J.; Rabl, P.; Moiseyev, N.; and Rotter, S. Dynamically encircling an exceptional point for asymmetric mode switching. {\em Nature} {\bf 2016}, {\em 537}, 76-80.
	
	\bibitem{Dym5} Hassan, A.U.; Zhen, B.; Solja${\check {\rm c}}$i\'c, M.; Khajavikhan, M.; and Christodoulides, D.N.; Dynamically Encircling Exceptional Points: Exact Evolution and Polarization State Conversion. {\em Phys. Rev. Lett.} {\bf 2017}, {\em 118}, 093002.
	
	\bibitem{Dym6} Zhang, X.-L.; Wang, S.; Hou, B.; and Chan, C.T. Dynamically Encircling Exceptional Points: In situ Control of Encircling Loops and the Role of the Starting Point. {\em Phys. Rev. X} {\bf 2018}, {\em 8}, 021066.
	
	\bibitem{ColdAtom1} Li, J.; Harter, A.K.; Liu, J.; Melo, L.d.; Joglekar, Y.N.; Luo, L. Observation of parity-time symmetry breaking transitions in a dissipative Floquet system of ultracold atoms. {\em  Nat. Commun.} {\bf 2019}, {\em 10}, 855.
	
	\bibitem{ColdAtom2} Gou, W.; Chen, T.; Xie, D.; Xiao, T.; Deng, T.-S.; Gadway, B.; Yi, W.; and Yan, B. Tunable Nonreciprocal Quantum Transport through a Dissipative Aharonov-Bohm Ring in Ultracold Atoms. {\em Phys. Rev. Lett.} {\bf 2020}, {\em 124}, 070402.
	
	\bibitem{Photonic1} Zeuner, J.M.; Rechtsman, M.C.; Plotnik, Y.; Lumer, Y.; Nolte, S.; Rudner, M.S.; Segev, M.; and Szameit, A. Observation of a Topological Transition in the Bulk of a Non-Hermitian System. {\em Phys. Rev. Lett.} {\bf 2015}, {\em 115}, 040402.
	
	\bibitem{Photonic2} Weimann, S.; Kremer, M.; Plotnik, Y.; Lumer, Y.; Nolte, S.; Makris, K.G.; Segev, M.; Rechtsman, M.C.; and Szameit, A. Topologically protected bound states in photonic parity-time-symmetric crystals. {\em Nature Mater.} {\bf 2016}, {\em 16}, 433.
	
	\bibitem{Photonic3} Wang, K.; Qiu, X.; Xiao, L.; Zhan, X.; Bian, Z.; Sanders, B.C.;	Yi, W.; and Xue, P. Observation of emergent momentum-time skyrmions in parity-time-symmetric non-unitary quench dynamics. {\em  Nat. Commun.} {\bf 2019}, {\em 10}, 2293.
	
	\bibitem{Photonic4} Xiao, L.; Deng, T.; Wang, K.; Zhu, G.; Wang, Z.; Yi, W.; and Xue, P. Non-Hermitian bulk-boundary correspondence in quantum dynamics. {\em Nat. Phys.} {\bf 2020}, https://doi.org/10.1038/s41567-020-0836-6.
	
	\bibitem{Acoustic1} Zhu, W.; Fang, X.; Li, D.; Sun, Y.; Li, Y.; Jing, Y.; and Chen, H.; Simultaneous Observation of a Topological Edge State and Exceptional Point in an Open and Non-Hermitian Acoustic System. {\em Phys. Rev. Lett.} {\bf 2018}, {\em 121}, 124501.
	
	\bibitem{Acoustic2} Shen, C.; Li, J.; Peng, X.; and Cummer, S.A. Synthetic exceptional points and unidirectional zero reflection in non-Hermitian acoustic systems. {\em Phys. Rev. Materials} {\bf 2018}, {\em 2}, 125203.
	
	\bibitem{Acoustic3} Gao, H.; Xue, H.; Wang, Q.; Gu, Z.; Liu, T.; Zhu, J.; and Zhang, B. Observation of topological edge states induced solely by non-Hermiticity in an acoustic crystal. {\em Phys. Rev. B} {\bf 2020}, {\em 101}, 180303(R).
	
	\bibitem{Circuit1} Hofmann, T.; Helbig, T.; Schindler, F.; Salgo, N.; Brzezi\'nska, M.; Greiter, M.; Kiessling, T.; Wolf, D.; Vollhardt, A.; Kaba${\check {\rm s}}$i, A.; Lee, C.H.; Bilu${\check {\rm s}}$i\'c, A.; Thomale, R.; and Neupert, T. Reciprocal skin effect and its realization in a topolectrical circuit. {\em Phys. Rev. Research} {\bf 2020}, {\em 2}, 023265.
	
	\bibitem{Circuit2} Helbig, T.; Hofmann, T.; Imhof, S.; Abdelghany, M.; Kiessling, T.; Molenkamp, L.W.; Lee, C.H.; Szameit, A.; Greiter, M.; and Thomale; R. Generalized bulk-boundary correspondence in non-Hermitian topolectrical circuits. {\em Nat. Phys.} {\bf 2020}, https://doi.org/10.1038/s41567-020-0922-9.
	
	\bibitem{Circuit3} Liu, S.; Ma, S.; Yang, C.; Zhang, L.; Gao, W.; Xiang, Y.J.; Cui, T.J.; and Zhang, S. Gain- and Loss-Induced Topological Insulating Phase in a Non-Hermitian Electrical Circuit. {\em Phys. Rev. Applied} {\bf 2020}, {\em 13}, 014047.
	
	\bibitem{NVCent1} Wu, Y.; Liu, W.; Geng, J.; Song, X.; Ye, X.; Duan, C.-K.; Rong, X.; Du, J. Observation of parity-time symmetry breaking in a single-spin system. {\em Science} {\bf 2019}, {\em 364}, 878-880.
	
	\bibitem{TopoLas1} Harari, G.; Bandres, M.A.; Lumer, Y.; Rechtsman, M.C.; Chong, Y.D.; Khajavikhan, M.; Christodoulides, D.N.; and Segev, M. Topological insulator laser: Theory. {\em Science} {\bf 2018}, {\em 359}, eaar4003.
	
	\bibitem{TopoLas2} Bandres, M.A.; Wittek, S.; Harari, G.; Parto, M.; Ren, J.; Segev, M.; Christodoulides, D.N.; and Khajavikhan, M.; Topological insulator laser: Experiments. {\em Science} {\bf 2018}, {\em 359}, eaar4005.
	
	\bibitem{TopoLas3} Kartashov, Y.V.; and Skryabin, D.V. Two-Dimensional Topological Polariton Laser. {\em Phys. Rev. Lett.} {\bf 2019}, {\em 122}, 083902.
	
	\bibitem{EPSense1} Wiersig, J. Enhancing the Sensitivity of Frequency and Energy Splitting Detection by Using Exceptional Points: Application to Microcavity Sensors for Single-Particle Detection. {\em Phys. Rev. Lett.} {\bf 2014}, {\em 112}, 203901.
	
	\bibitem{EPSense2} Lau, H.-K.; and Clerk, A.A. Fundamental limits and non-reciprocal approaches in non-Hermitian quantum sensing. {\em Nat. Commun.} {\bf 2018}, {\em 9}, 4320.
	
	\bibitem{EPSense3} Hodaei, H.; Hassan, A.U.; Wittek, S.; Garcia-Gracia, H.; ElGanainy, R.; Christodoulides, D.N.; and Khajavikhan, M. Enhanced sensitivity at higher-order exceptional points. {\em Nature} {\bf 2017}, {\em 548}, 187-191.
	
	\bibitem{EPSense4} Chen, W.; \"Ozdemir, S.K.; Zhao, G.; Wiersig, J.; and Yang, L. {\em Nature} {\bf 2017}, {\em 548}, 192-196.
	
	\bibitem{ZhouNHFTP1} Zhou, L.; and Gong, J. Non-Hermitian Floquet topological phases with arbitrarily many real-quasienergy edge states. {\em Phys. Rev. B} {\bf 2018}, {\em 98}, 205417.
	
	\bibitem{ZhouNHFTP2} Zhou, L.; and Pan, J. Non-Hermitian Floquet topological phases in the double-kicked rotor. {\em Phys. Rev. A} {\bf 2019}, {\em 100}, 053608.
	
	\bibitem{ZhouNHFTP3} Zhou, L. Dynamical characterization of non-Hermitian Floquet topological phases in one dimension. {\em Phys. Rev. B} {\bf 2019}, {\em 100}, 184314.
	
	\bibitem{ZhouNHFTP4} Zhou, L. Non-Hermitian Floquet topological superconductors with multiple Majorana edge modes. {\em Phys. Rev. B} {\bf 2020}, {\em 101}, 014306.
	
	\bibitem{ZhouNHFTP5} Pan, J.; and Zhou, L. Non-Hermitian Floquet second order topological insulators in periodically quenched lattices. {\em arXiv:2004.06283} {\bf 2020}.
	
	\bibitem{NHFTP1} Yuce, C. PT symmetric Floquet topological phase. {\em Eur. Phys. J. D} {\bf 2015}, {\em 69}, 184.
	
	\bibitem{NHFTP2} Turker, Z.; Tombuloglu, S.; and Yuce, C. PT symmetric Floquet topological phase in SSH model. {\em Phys. Lett. A} {\bf 2018}, {\em 382}, 2013-2016.
	
	\bibitem{NHFTP3} Li, M.; Ni, X.; Weiner, M.; Al\`u, A.; and Khanikaev, A.B. Topological phases and nonreciprocal edge states in non-Hermitian Floquet insulators. {\em Phys. Rev. B} {\bf 2019}, {\em 100}, 045423.
	
	\bibitem{NHFTP4} Zhang, X.; and Gong, J. Non-Hermitian Floquet topological phases: Exceptional points, coalescent edge modes, and the skin effect. {\em Phys. Rev. B} {\bf 2020}, {\em 101}, 045415.
	
	\bibitem{NHFTP5} Longhi, S. Floquet exceptional points and chirality in non-Hermitian Hamiltonians. {\em J. Phys. A: Math. Theor.} {\bf 2017}, {\em 50}, 505201.
	
	\bibitem{NHFTP6} Chitsazi, M.; Li, H.; Ellis, F.M.; and Kottos, T. Experimental Realization of Floquet ${\cal PT}$-Symmetric Systems. {\em Phys. Rev. Lett.} {\bf 2017}, {\em 119}, 093901.
	
	\bibitem{NHFTP7} Le\'on-Montiel, R.d.J.; Quiroz-Ju\'arez, M.A.; Dom\'inguez-Ju\'arez, Quintero-Torres, J.L.R.; Arag\'on, J.L.; Harter, A.K.; Joglekar, Y.N. Observation of slowly decaying eigenmodes without exceptional points in Floquet dissipative synthetic circuits. {\em Communications Physics} {\bf 2018}, {\em 1}, 88.
	
	\bibitem{NHFTP8} Lee, C.H.; and Longhi, S. Ultrafast and Anharmonic Rabi Oscillations between Non-Bloch-Bands. {\em arXiv:2003.10763} {\bf 2020}.
	
	\bibitem{NHFTP9} Wu H.; and An, J.-H. Floquet Topological Phases of Non-Hermitian Disordered Systems. {\em arXiv:2003.08055} {\bf 2020}.
	
	\bibitem{NHFTP10} He, P.; and Huang, Z.-H. Floquet-engineering and simulating exceptional rings with a quantum spin system. {\em arXiv:2005.02703} {\bf 2020}.
	
	\bibitem{CL0} Creutz, L. End States, Ladder Compounds, and Domain-Wall Fermions. {\em Phys. Rev. Lett.} {\bf 1999}, {\em 83}, 2636.
	
	\bibitem{CLExp1} Kremer, M.; Petrides, I.; Meyer, E.; Heinrich, M.; Zilberberg, O.; and Szameit, A. A square-root topological insulator with non-quantized indices realized with photonic Aharonov-Bohm cages. {\em Nat. Commun.} {\bf 2020}, {\em 11}, 907.
	
	\bibitem{CLExp2} Mukherjee, S.; Liberto, M.D.; \"Ohberg, P.; Thomson, R.R.; and Goldman, N. Experimental Observation of Aharonov-Bohm Cages in Photonic Lattices. {\em Phys. Rev. Lett.} {\bf 2018}, {\em 121}, 075502.
	
	\bibitem{CLExp3} Kang, J.H.; Han, J.H.; and Shin, Y. Creutz ladder in a resonantly shaken 1D optical lattice. {\em New J. Phys.} {\bf 2020}, {\em 22}, 013023.
	
	\bibitem{CLExp4} Li, X.; Zhao, E.; and Liu, W.V. Topological states in a ladder-like optical lattice containing ultracold atoms in higher orbital bands. {\em Nat. Commun.} {\bf 2013}, {\em 4}, 1523.
	
	\bibitem{CL1} Gligori\'c, G.; Leykam, D.; and Maluckov A.; Influence of different disorder types on Aharonov-Bohm caging in the diamond chain. {\em Phys. Rev. A} {\bf 2020}, {\em 101}, 023839.
	
	\bibitem{CL2} Liberto, M.D.; Mukherjee, S.; and Goldman, N. Nonlinear dynamics of Aharonov-Bohm cages. {\em Phys. Rev. A} {\bf 2019}, {\em 100}, 043829.
	
	\bibitem{CL3} Sun N.; and Lim, L.-K. Quantum charge pumps with topological phases in a Creutz ladder. {\em Phys. Rev. B} {\bf 2017}, {\em 96}, 035139.
	
	\bibitem{CL4} Kuno, Y.; Orito, T.; and Ichinose, I. Flat-band many-body localization and ergodicity breaking in the Creutz ladder. {\em New J. Phys.} {\bf 2020}, {\em 22}, 013032.
	
	\bibitem{CL5} Kuno, Y. Extended flat-bands, entanglement and topological properties in a Creutz ladder. {\em Phys. Rev. B} {\bf 2020}, {\em 101}, 184112.
	
	\bibitem{CL6} Zurita, J.; Creffield, C.E.; and Platero, G. Topology and Interactions in the Photonic Creutz and Creutz-Hubbard Ladders. {\em Adv. Quantum Technol.} {\bf 2019}, {\em 3}, 1900105.
	
	\bibitem{CL7} Sticlet, D.; Seabra, L.; Pollmann, F.; and Cayssol, J. From fractionally charged solitons to Majorana bound states in a one-dimensional interacting model. {\em Phys. Rev. B} {\bf 2014}, {\em 89}, 115430.
	
	\bibitem{CL8} J\"unemann, J.; Piga, A.; Ran, S.-J.; Lewenstein, M.; Rizzi, M.; and Bermudez, A. Exploring Interacting Topological Insulators with Ultracold Atoms: The Synthetic Creutz-Hubbard Model. {\em Phys. Rev. X} {\bf 2017}, {\em 7}, 031057.
	
	\bibitem{CL9} Yang, F.; Perrin, V.; Petrescu, A.; Garate, I.; and Hur, K.L. From topological superconductivity to quantum Hall states in coupled wires. {\em Phys. Rev. B} {\bf 2020}, {\em 101}, 085116.
	
	\bibitem{CL10} Haller, A.; Rizzi, M.; and Filippone, M. Drude weight increase by orbital and repulsive interactions in fermionic ladders. {\em Phys. Rev. Research} {\bf 2020}, {\em 2}, 023058.
	
	\bibitem{SCL1} Santos, R.A.; and B\'eri, B. Fractional topological insulator precursors in spin-orbit fermion ladders. {\em Phys. Rev. B} {\bf 2019}, {\em 100}, 235122.

	\bibitem{SCL2} Het\'enyi, B.; and Yahyavi, M. Topological insulation in a ladder model with particle-hole and reflection symmetries. {\em  J. Phys.: Condens. Matter} {\bf 2018}, {\em 30}, 10LT01.
	
	\bibitem{SCL3} Gholizadeh, S.; Yahyavi, M.; and Het\'enyi, B. Extended Creutz ladder with spin-orbit coupling: A one-dimensional analog of the Kane-Mele model. {\em  Europhys. Lett.} {\bf 2018}, {\em 122}, 27001.
	
	\bibitem{FSCL} Zhou, L.; and Du, Q.; Floquet topological phases with fourfold-degenerate edge modes in a driven spin-1/2 Creutz ladder. {\em  Phys. Rev. A} {\bf 2020}, {\em 101}, 033607.
	
	\bibitem{AsbothSTF1} Asb\'oth, J.K. Symmetries, topological phases, and bound states in the one-dimensional quantum walk. {\em  Phys. Rev. B} {\bf 2012}, {\em 86}, 195414.
	
	\bibitem{AsbothSTF2} Asb\'oth, J.K.; and Obuse, H. Bulk-boundary correspondence for chiral symmetric quantum walks. {\em  Phys. Rev. B} {\bf 2013}, {\em 88}, 121406(R).
	
	\bibitem{STPTab} Ryu, S.; Schnyder, A.P.; Furusaki, A.; and Ludwig, A.W.W. Topological insulators and superconductors: tenfold way and dimensional hierarchy. {\em  New J. Phys.} {\bf 2010}, {\em 12}, 065010.
	
	\bibitem{FTPTab} Roy R.; and Harper, F. Periodic table for Floquet topological insulators. {\em  Phys. Rev. B} {\bf 2017}, {\em 96}, 155118.
	
	\bibitem{NHSkin} Yao S.; and Wang, Z. Edge States and Topological Invariants of Non-Hermitian Systems. {\em  Phys. Rev. Lett.} {\bf 2018}, {\em 121}, 086803.
	
	\bibitem{SKR} Zhou, L.; and Gong, J. Floquet topological phases in a spin-1/2 double kicked rotor. {\em  Phys. Rev. A} {\bf 2018}, {\em 97}, 063603.
	
	\bibitem{QMatrix} Song, F.; Yao, S.; and Wang, Z. Non-Hermitian Topological Invariants in Real Space. {\em  Phys. Rev. Lett.} {\bf 2019}, {\em 123}, 246801.
	
	\bibitem{BBC1} Shapiro, J. The bulk-edge correspondence in three simple cases. {\em Reviews in Mathematical Physics} {\bf 2020}, {\em 32}, 2030003.
	
	\bibitem{QMatrix2} Mondragon-Shem, I.; Hughes, T.L.; Song, J.; and Prodan, E. Topological Criticality in the Chiral-Symmetric AIII Class at Strong Disorder. {\em  Phys. Rev. Lett.} {\bf 2014}, {\em 113}, 046802.
	
	\bibitem{QMatrix3} Song J.; and Prodan, E. AIII and BDI topological systems at strong disorder. {\em  Phys. Rev. B} {\bf 2014}, {\em 89}, 224203.
	
	\bibitem{MCD1} Cardano, F.; D'Errico, A.; Dauphin, A.; Maffei, M.; Piccirillo, B.;	Lisio, C.d.; Filippis, G.D.; Cataudella, V.; Santamato, E.; Marrucci, L.; Lewenstein, M.; and Massignan, P. Detection of Zak phases and topological invariants in a chiral quantum walk of twisted photons. {\em Nat. Commun.} {\bf 2017}, {\em 8}, 15516.
	
	\bibitem{MCD2} Maffei, M.; Dauphin, A.; Cardano, F.; Lewenstein, M.; and Massignan, P.; Topological characterization of chiral models through their long time dynamics.  {\em New J. Phys.} {\bf 2018}, {\em 20}, 013023.
	
	\bibitem{MCD3} Bomantara, R.W.; Zhou, L.; Pan, J.; and Gong, J. Coupled-wire construction of static and Floquet second-order topological insulators. {\em  Phys. Rev. B} {\bf 2019}, {\em 99}, 045441.
	
	\bibitem{MCD4} Haller, A.; Massignan, P.; Rizzi M. Detecting topology through dynamics in interacting fermionic wires. {\em arXiv:2001.09074} {\bf 2020}.
	
	\bibitem{MCD5} D'Errico, A.; Colandrea, F.D.; Barboza, R.; Dauphin, A.; Lewenstein, M.;  Massignan, P.; Marrucci, L.; and Cardano, F. Bulk detection of time-dependent topological transitions in quenched chiral models. {\em  Phys. Rev. Research} {\bf 2020}, {\em 2}, 023119.
	
	\bibitem{MCD6} Meier, E.J.; An, F.A.; Dauphin, A.; Maffei, M.; Massignan, P.; Hughes, T.L.; and Gadway, B. Observation of the topological Anderson insulator in disordered atomic wires. {\em Science} {\bf 2018}, {\em 362}, 929.
	
	\bibitem{MCD7} Xie, D.; Deng, T.-S.; Xiao, T.; Gou, W.; Chen, T.; Yi, W.; and Yan, B. Topological Quantum Walks in Momentum Space with a Bose-Einstein Condensate. {\em  Phys. Rev. Lett.} {\bf 2020}, {\em 124}, 050502.

\end{thebibliography}


\end{document}